\documentclass[twocolumn]{aastex631}

\shorttitle{Structure and Evolution of AGN}

\shortauthors{Ananna et al.}
\graphicspath{{./}{}}
\usepackage{natbib}
\usepackage{amsmath}
\defcitealias{Ricci2017Nat}{R17b}
\defcitealias{Ananna2022}{A22}

\newcommand{\erdflogxistarunabs}{-4.29}
\newcommand{\erdfdeltaaunabs}{$0.10^{+0.16}_{-0.00}$}
\newcommand{\erdfepislonlamunabs}{$2.73^{+0.22}_{-0.16}$}
\newcommand{\erdfloglamstarunabs}{$-1.064^{+0.055}_{-0.052}$}

\newcommand{\erdflogxistarLogNHTwentyTwoToTwentyFive}{-3.42}
\newcommand{\erdfdeltaaLogNHTwentyTwoToTwentyFive}{$-0.01^{+0.27}_{-0.34}$}
\newcommand{\erdfepislonlamLogNHTwentyTwoToTwentyFive}{$2.51^{+0.32}_{-0.19}$}
\newcommand{\erdfloglamstarLogNHTwentyTwoToTwentyFive}{$-1.750\pm0.093$}

\newcommand{\erdflogxistarunabssigfive}{-4.26}
\newcommand{\erdfdeltaaunabssigfive}{$0.10^{+0.55}_{-0.05}$}
\newcommand{\erdfepislonlamunabssigfive}{$2.51^{+0.44}_{-0.46}$}
\newcommand{\erdfloglamstarunabssigfive}{$-1.165^{+0.112}_{-0.076}$}

\newcommand{\erdflogxistarLogNHTwentyTwoToTwentyFivesigfive}{-3.42}
\newcommand{\erdfdeltaaLogNHTwentyTwoToTwentyFivesigfive}{$0.03^{+0.30}_{-0.65}$}
\newcommand{\erdfepislonlamLogNHTwentyTwoToTwentyFivesigfive}{$2.91^{+0.38}_{-0.35}$}
\newcommand{\erdfloglamstarLogNHTwentyTwoToTwentyFivesigfive}{$-1.67^{+0.13}_{-0.14}$}

\begin{document}

% \title{BASS [NUMBER]: Obscuration and Eddington Ratio of Rapidly Growing Supermassive Black Holes} %variable structure of the torus

\title{Probing the Structure and Evolution of BASS AGN through Eddington Ratios} %variable structure of the torus

\author[0000-0001-8211-3807]{Tonima Tasnim Ananna}
\affiliation{Department of Physics and Astronomy, Dartmouth College, 6127 Wilder Laboratory, Hanover, NH 03755, USA}

\author[0000-0002-0745-9792]{C. Megan Urry}
%\affiliation{Yale Center for Astronomy \& Astrophysics, Physics Department, New Haven, CT 06520, USA}
\affiliation{Department of Physics and Yale Center for Astronomy \& Astrophysics, Yale University, P.O. Box 201820, New Haven, CT 06520, USA}

\author[0000-0001-5231-2645]{Claudio Ricci}
\affiliation{N\'ucleo de Astronom\'ia de la Facultad de Ingenier\'ia, Universidad Diego Portales, Av. Ej\'ercito Libertador 441, Santiago 22, Chile}
\affiliation{Kavli Institute for Astronomy and Astrophysics, Peking University, Beijing 100871, People's Republic of China}

\author[0000-0002-5554-8896]{Priyamvada Natarajan}
\affiliation{Department of Astronomy, Yale University, 52 Hillhouse Avenue, New Haven, CT 06520, USA}
\affiliation{Department of Physics, Yale University, P.O. Box 208121, New Haven, CT 06520, USA}
\affiliation{Black Hole Initiative, Harvard University, 20 Garden Street, Cambridge MA 02138, USA}

\author[0000-0003-1468-9526]{Ryan C. Hickox}
\affiliation{Department of Physics and Astronomy, Dartmouth College, 6127 Wilder Laboratory, Hanover, NH 03755, USA}

\author[0000-0002-3683-7297]{Benny Trakhtenbrot}
\affiliation{School of Physics and Astronomy, Tel Aviv University, Tel Aviv 69978, Israel}

\author[0000-0001-7568-6412]{Ezequiel Treister}
\affil{Instituto de Astrof{\'{\i}}sica, Facultad de F{\'{i}}sica, Pontificia Universidad Cat{\'{o}}lica de Chile, Campus San Joaquín, Av. Vicuña Mackenna 4860, Macul Santiago, Chile, 7820436} 

\author[0000-0002-5489-4316]{Anna K. Weigel}
\affiliation{Modulos AG, Technoparkstrasse 1, CH-8005 Zurich, Switzerland}
% \affiliation{Department of Physics, ETH Zurich, Wolfgang-Pauli-Strasse 27, CH-8093 Zurich, Switzerland}

%\collaboration{6}{(AAS Journals Data Editors)}

\author[0000-0001-7821-6715]{Yoshihiro Ueda}
\affiliation{Department of Astronomy, Kyoto University, Kitashirakawa-Oiwake-cho, Sakyo-ku, Kyoto 606-8502, Japan}

\author[0000-0002-7998-9581]{Michael J. Koss}
\affiliation{Eureka Scientific, 2452 Delmer Street Suite 100, Oakland, CA 94602-3017, USA}
\affiliation{Space Science Institute, 4750 Walnut Street, Suite 205, Boulder, Colorado 80301, USA}

\author[0000-0002-8686-8737]{F. E. Bauer}
\affil{Instituto de Astrof{\'{\i}}sica, Facultad de F{\'{i}}sica, Pontificia Universidad Cat{\'{o}}lica de Chile, Campus San Joaquín, Av. Vicuña Mackenna 4860, Macul Santiago, Chile, 7820436} 
\affil{Centro de Astroingenier{\'{\i}}a, Facultad de F{\'{i}}sica, Pontificia Universidad Cat{\'{o}}lica de Chile, Campus San Joaquín, Av. Vicuña Mackenna 4860, Macul Santiago, Chile, 7820436} 
\affil{Millennium Institute of Astrophysics, Nuncio Monse{\~{n}}or S{\'{o}}tero Sanz 100, Of 104, Providencia, Santiago, Chile} 

\author[0000-0001-8433-550X]{Matthew J. Temple}
\affil{N\'ucleo de Astronom\'ia de la Facultad de Ingenier\'ia, Universidad Diego Portales, Av. Ej\'ercito Libertador 441, Santiago 22, Chile}

\author[0000-0003-0476-6647]{Mislav Balokovi\'{c}}
\affiliation{Department of Physics and Yale Center for Astronomy \& Astrophysics, Yale University, P.O. Box 201820, New Haven, CT 06520, USA}
\affiliation{Department of Astronomy, Yale University, 52 Hillhouse Avenue, New Haven, CT 06520, USA}

\author[0000-0002-7962-5446]{Richard Mushotzky}
\affiliation{Department of Astronomy, University of Maryland, College Park, MD 20742, USA}

\author[0000-0002-5504-8752]{Connor Auge}
\affiliation{Institute for Astronomy, University of Hawai\`{}i, 2680 Woodlawn Drive, Honolulu, HI 96822, USA}

\author[0000-0002-1233-9998]{David B. Sanders}
\affiliation{Institute for Astronomy, University of Hawai\`{}i, 2680 Woodlawn Drive, Honolulu, HI 96822, USA}

\author[0000-0002-2603-2639]{Darshan Kakkad}
\affiliation{Space Telescope Science Institute, 3700 San Martin Drive, Baltimore, 21218 MD, USA}

\author[0000-0001-8020-3884]{Lia F. Sartori}
\affiliation{ETH Zurich, Institute for Particle Physics and Astrophysics, Wolfgang-Pauli-Strasse 27, CH-8093 Zurich, Switzerland}

\author[0000-0001-5544-0749]{Stefano Marchesi}
\affiliation{INAF - Osservatorio di Astrofisica e Scienza dello Spazio di Bologna, Via Piero Gobetti, 93/3, 40129, Bologna, Italy}
\affiliation{Department of Physics and Astronomy, Clemson University, Kinard Lab of Physics, Clemson, SC 29634, USA}

\author{Fiona Harrison}
\affiliation{Cahill Center for Astronomy and Astrophysics, California Institute of Technology, Pasadena, CA 91125, USA}

\author[0000-0003-2686-9241]{Daniel Stern}
\affiliation{Jet Propulsion Laboratory, California Institute of Technology, 4800 Oak Grove Drive, MS 169-224, Pasadena, CA 91109, USA}

\author[0000-0002-5037-951X]{Kyuseok Oh}
\affiliation{Korea Astronomy \& Space Science institute, 776, Daedeokdae-ro, Yuseong-gu, Daejeon 34055, Republic of Korea}
\affiliation{Department of Astronomy, Kyoto University, Kitashirakawa-Oiwake-cho, Sakyo-ku, Kyoto 606-8502, Japan}

\author[0000-0002-9144-2255]{Turgay Caglar}
\affiliation{Leiden Observatory, PO Box 9513, 2300 RA, Leiden, the Netherlands}

\author[0000-0003-2284-8603]{Meredith C. Powell}
\affiliation{Kavli Institute of Particle Astrophysics and Cosmology, Stanford University, 452 Lomita Mall, Stanford, CA 94305, USA}

\author[0000-0003-1468-9526]{Stephanie A. Podjed}
\affiliation{Department of Physics and Astronomy, Dartmouth College, 6127 Wilder Laboratory, Hanover, NH 03755, USA}

\author[0000-0001-8450-7463]{Julian E. Mej\'ia-Restrepo}
\affiliation{European Southern Observatory, Casilla 19001, Santiago 19, Chile}

\newcommand{\lamEdd }{\ifmmode \lambda_{\rm Edd} \else $\lambda_{\rm Edd}$\fi}
\newcommand{\lamEddeff}{\ifmmode \lambda_{\rm eff} \else $\lambda_{\rm eff}$\fi}
\newcommand{\NH }{\ifmmode N_{\rm H} \else $N_{\rm H}$\fi}
\newcommand{\logNH }{\ifmmode \log (N_{\rm H}/{\rm cm}^{-2}) \else $\log (N_{\rm H}/{\rm cm}^{-2})$\fi}	
\newcommand{\Mbh  }{\ifmmode M_{\rm BH} \else $M_{\rm BH}$\fi}
\newcommand{\Msun}{\ifmmode M_{\odot} \else $M_{\odot}$\fi}



\begin{abstract}

%\textbf
{We constrain the intrinsic Eddington ratio (\lamEdd ) distribution function for local AGN in bins of low and high obscuration [$\logNH \leq 22$ and $22 < \logNH < 25$], using the Swift-BAT 70-month/BASS DR2 survey. We interpret the fraction of obscured AGN in terms of circum-nuclear geometry and temporal evolution. Specifically, at low Eddington ratios ($\log~\lamEdd < -2$), obscured AGN outnumber unobscured ones by a factor of $\sim4$, reflecting the covering factor of the circum-nuclear material (0.8, or a torus opening angle of $\sim34^\circ$). At high Eddington ratios ($\log~\lamEdd > -1$), the trend is reversed, with $< 30$\% of AGN having $\logNH > 22$, which we suggest is mainly due to the small fraction of time spent in a highly obscured state.
Considering the Eddington ratio distribution function of narrow-line and broad-line AGN from our prior work, we see a qualitatively similar picture. To disentangle temporal and geometric effects at high \lamEdd, we explore plausible clearing scenarios such that the time-weighted covering factors agree with the observed population ratio. We find that the low fraction of obscured AGN at high \lamEdd\ is primarily due to the fact that the covering factor drops very rapidly, with more than half the time %at those \lamEdd\ is 
spent with $< 10\%$ covering factor. We also find that nearly all obscured AGN at high-\lamEdd\ exhibit some broad-lines. We suggest that this is because the height of the depleted torus falls below the height of the broad-line region, making the latter visible from all lines of sight.}%, and a final remaining covering factor $<5$\%.}

% We constrain the intrinsic Eddington ratio (\lamEdd ) distribution function for local AGN in bins of obscuration [$\logNH \leq 20$ and $22 < \logNH < 25$], using the Swift-BAT 70-month/BASS DR2 survey. We compare our results with the Eddington ratio distribution function of broad-line and narrow-line AGN, calculated in our prior work, and  interpret these differences in terms of their circum-nuclear geometry and temporal evolution. We find that at low Eddington ratios ($\log~\lamEdd < -2$), obscured AGN outnumber unobscured ones by a factor of $\sim5$, reflecting the geometry of the circum-nuclear material. At high Eddington ratios ($\log~\lamEdd > -1$), the trend is reversed, with $< 30$\% of AGN having $\logNH > 22$, and less than 5\% being narrow-line AGN. Infrared studies find heavily obscured, broad-line AGN at high luminosities/\lamEdd\ (e.g., HotDOGs and red quasars). Our analysis of intrinsic Eddington ratio distribution functions also suggest that at high \lamEdd , most $\logNH > 22$ AGN are broad-line, as narrow-line AGN space densities fall rapidly with increasing \lamEdd . As obscured, high-\lamEdd\ AGN tend to have high covering factors, we suggest that the low fraction of obscured AGN at high \lamEdd\ reveals the relative longevity of the broad-line obscured phase (15--25\%), rather than the geometric covering factor. We also present the covering factors/angles for the full torus, and for the region of the torus that blocks out the broad-line region completely.

\end{abstract}


%% and \citet commands to identify citations. The citations are

\section{Introduction} \label{sec:intro}

To study active galactic nuclei (AGN) comprehensively, across multiple wavelengths, we must disentangle confounding observational selection effects, to measure the underlying physical quantities like AGN mass, luminosity, and accretion rate.
For example, although unobscured AGN dominate optically-selected samples of highly luminous sources \citep[e.g.,][]{Richards2002} they end up being a minority in the census of the total population \citep[e.g.,][]{treister2004, gilli2007, Treister2009, ananna2019}. %(Treister et al. 2004, Treister, Virani and Urry 2009, SWIFT BAT papers, Ananna et al. 2019, other REFS).
Similarly, radio-selected samples are dominated by radio-loud AGN, which also constitutes a very small fraction of the overall AGN population \citep[e.g.,][]{best2005,stawarz2010}. To get at the underlying demographics, it is important to account for both selection biases and measurement uncertainties. In the last fifteen years, high-energy X-ray telescopes such as Swift-BAT, INTEGRAL and NuSTAR have provided nearly unbiased X-ray-detected samples of AGN, where biases only become significant for the most heavily obscured sources \citep[see][]{Ricci:2015aa,koss2016ctk, Ananna2022}. The Swift-BAT 70--month survey provided the most sensitive map above 10 keV \citep{Baumgartner:2013aa}. Using optical and infrared spectroscopy, the BAT AGN spectroscopic survey (BASS) provided morphology, masses, redshifts, luminosity and obscuring column density for 752 non-blazar AGN, including 292 Type~2 AGN \citep{Ricci2017_Xray_cat,koss2022_agn_catalog,koss2022_DR2_data}. The 98\% completeness in black hole mass estimates for unbeamed AGN outside of the Galactic plane comes from broad emission lines and from velocity dispersion of stars within the host galaxy bulges \citep{Mejia_DR2_BLR,koss2022_DR2_data,koss2022_host_galaxy_vel_disp}.
 
\begin{figure}[ht!]
    \includegraphics[width=0.5\textwidth]{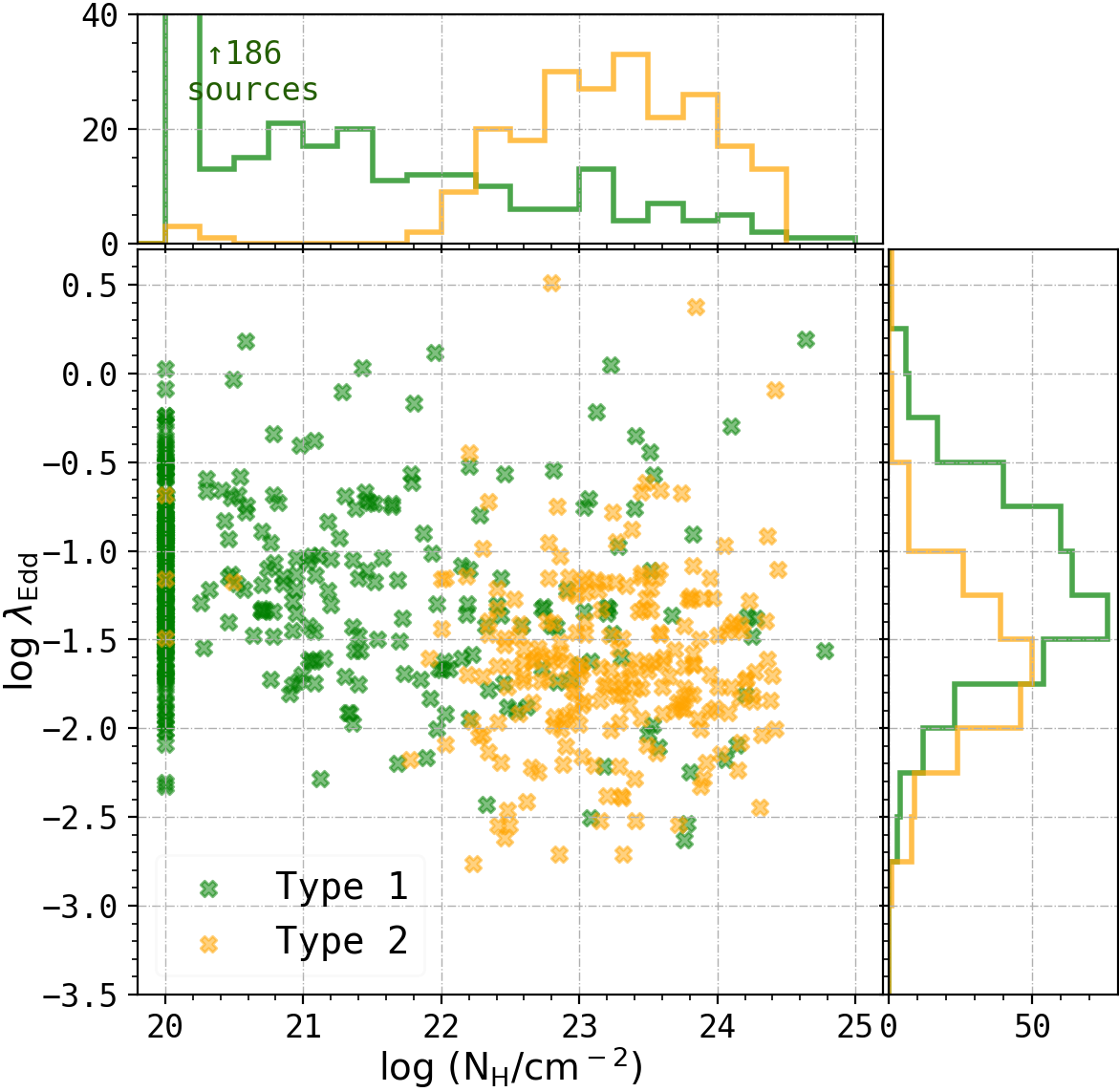}
	\caption{\label{fig:2d_scatter_plot} %\textit{Top panel:}
	The distribution of Type~1 (green crosses) and Type~2 (orange crosses) AGN in column density-Eddington ratio space. 
	There are 276 obscured AGN ($\logNH \geq 22$), compared to 220 Type~2 AGN (narrow lines). 
	%The number of obscured AGN with $\logNH \geq 22$ is 276, whereas Type~2 AGN is 220. 
	Similarly, there are 301 AGN with $\logNH < 22$, and 366 Type~1 AGN, % many Type~1 AGN showing
	some of which have high column densities.} 
	%\textit{Bottom panel:} Eddington ratio distribution functions for unobscured [\logNH $= 20{-}22$; blue line] and obscured [\logNH $= 22{-}25$; red line] AGN, along with the {A22} ERDFs for %broad-line/
	%Type~1 (green line) and %narrow-line/
	%Type~2 AGN (orange line) from {A22}. 
	%The ratio of obscured to unobscured is 4:1 at low Eddington ratio ($\log \lamEdd <-2$) and inverts at high values ($\log \lamEdd >-1$). 
	%Shaded regions show the 1 $\sigma$ uncertainties.}
	%categorization. In this Figure (as well as all the ERDFs shown in this work), the shaded region shows 1 $\sigma$ uncertainty on these function. %For comparison, we also show the ERDFs of Type~1 and Type~2 AGN calculated in {A22}, based on optical broad-line/narrow-line 
	%The black vertical lines represent effective Eddington ratios for different densities of gas according to \citet{fabian_effect_2008}.
	 % caption
\end{figure}

In \citeauthor{Ananna2022} (\citeyear{Ananna2022}; henceforth {A22}) we used a Bayesian inference methodology (described in detail in \S3 of {A22}, summarized here in \S\ref{sec:analysis}) to calculate the bias-corrected intrinsic black hole mass function (BHMF) and Eddington ratio distribution function (ERDF) of local AGN (i.e., $0.01 \leq z \leq 0.3$), divided into optical broad-line/Type~1 and narrow-line/Type~2 AGN categories. The bias correction accounted for both Eddington bias and the effect of obscuration on apparent source brightness (and thus, obscuration-dependent survey depth). {A22} found the shape of the BHMFs of Type~1 and Type~2 AGN to be in agreement. However, the distributions of Eddington ratios of Type~1 and Type~2 AGN were significantly different, and these differences prompt an interesting interpretation with far-reaching implications for AGN unification. %, which we present and discuss in this paper. %{A22} also divided the sample into two different mass bins ($\log~\Mbh < 10^8~\Msun$, $\log~\Mbh > 10^8~\Msun$) and reported that the \textit{shape of the ERDF is mass independent}. 
%In this work, we use the same methodology for further analysis of the sample in bins of obscuration. %These results describes the observational bias corrected true space density of local AGN.
%It also tells us a lot about the structure and evolution of AGN. 

The original AGN unification scheme was purely geometric: viewing angle explained the major distinctions amongst different types of AGN. 
However, AGN are variable sources, sometimes changing from one type to another \citep[e.g.][]{urry1995,marchese_ngc_2012,braito_ngc_2014,trakhtenbrot_1es_2019,green_time_2022}. 
%urry1995,bianchi_search_2005,marchese_ngc_2012,braito_ngc_2014,denney_typecasting_2014,lamassa_discovery_2015, guo_importance_2016,ricci_ic_2016,runnoe_now_2016,lamassa_insight_2017,noda_explaining_2018,yang_discovery_2018,macleod_changing-look_2019,trakhtenbrot_1es_2019,green_time_2022
%Luminous AGN, in particular, are likely to disrupt gas and dust near the central engine \citep[e.g.,][]{lawerence1991}. AGN properties also depend on black hole mass, accretion rate, as well as host galaxy properties. 
A luminosity-dependent interpretation referred to as the receding torus model \citep[e.g.,][]{lawerence1991,simpson2005,kyuseok2015} suggested that as AGN luminosity increases, since the dust sublimation radius increases, this causes the inner edge of the obscuring torus to recede. This idea was supported by the results of \citet{Ueda:2003aa}, \citet{lafranca2005}, \citet{barger2005}, \citet{simpson2005}, \citet{Treister2008}, who found a decreasing fraction of obscured AGN with luminosity. 

Analysis of the BASS sample supports an alternate interpretation, %which also has basis on theory \citep[e.g.,][]{fabian2006effedd,fabian_effect_2008}, 
where Eddington ratio rather than luminosity is the parameter that regulates how obscuring matter is distributed around AGN. \citet{Ricci2017Nat} used the observed BAT sample to study fractions of obscured AGN in two luminosity bins as a function of Eddington ratio. The obscured fractions in the two bins were very similar, and decreased with increasing Eddington ratio, implying that the torus structure is more fundamentally dependent on Eddington ratio than on luminosity. 
%As Eddington ratio takes into account both radiation pressure and gravitational pull, it could indeed be more intrinsically tied to the regulation of obscuring matter around AGN. 
%{R17b} also reported that when divided into bins on luminosity, the fraction of obscured AGN as a function of Eddington ratio stays essentially the same, which means \textit{Eddington ratio, rather the luminosity, regulates the shape of the torus}. 
This interpretation is known as the radiation-regulated unification model. The intrinsic Eddington ratio distribution function analysis for broad-line/narrow-line AGN in {A22} is also in agreement with these results. % (see Figure~1 from {R17b} to see the observed obscured fraction - \lamEdd\ relationship). 
\citet{ricci2022}, which is a companion paper to our analysis in this work, uses the updated BASS DR2 data \citep{koss2022_agn_catalog,koss2022_DR2_data} to reaffirm this version of the unification scheme, and discusses evolution along the obscuration${-}$Eddington ratio plane. 
Here, we report the intrinsic space density of AGN as a function of Eddington ratio in bins of obscuration, quantified by equivalent hydrogen column density, \logNH . Specifically, we divide the AGN into sub-samples according to the \logNH : %Specifically, we used four bins in absorption: 
$\logNH < 22$ and $22 \leq \logNH < 25$. Intrinsic ERDFs constrained in the X-ray-based column density measurements allow us to study this AGN sample from a different perspective and compare with the ERDFs derived earlier by {A22} using Type~1/Type~2 (optical broad-line/narrow-line) categorization.
%are more precise, and 
Moreover, X-ray-based column densities provide direct insight into line-of-sight circum-nuclear obscuration, and facilitate comparison with theoretical predictions. Figure~\ref{fig:2d_scatter_plot} shows the distribution of observed Type~1 and Type~2 AGN in \lamEdd -$\log~{\rm N}_{\rm H}$ space. %\textcolor{red}{Comment on observed distribution} 
The number of objects in each obscuration bin is given in Table~\ref{tab:_mcmc}. %We compare these results to the broad-line/narrow-line Eddington ratio distributions. We discuss the implications of our these intrinsic \lamEdd\ distributions in terms of torus structure and AGN evolution. %compare with the results for optical Type~1/Type~2 categorization (results deiscussed in next paragraph)

Our paper is structured as follows: in \S\ref{sec:analysis}, we discuss the data and the analysis methodology. In \S\ref{sec:result_only}, we present our results, and in \S~\ref{sec:discussion_only} we offer a physically motivated interpretation of our results, divided into high- and low-\lamEdd\ regimes, in the context of geometric unification and transition. In \S~\ref{sec:conclusion}, we present our conclusions. % in terms of geometry and spatial distribution of obscuring matter, and evolution of this geometry in different Eddington ratio regimes. %discuss our findings in the context of AGN structure and evolution.
%Specifically, in \S~\ref{sec:fabian_model} we compare our interpretation to prior theoretical work; in \S~\ref{sec:claudio} we discuss our results in the context of the radiation-regulated unification model; in \S~\ref{sec:glikman}, we discuss it in the context of other observational results and provide additional observational tests of our model; in \S~\ref{sec:timescale} we provide an estimate of time spent in obscured phase. 
We adopt a $\Lambda$CDM cosmology with $h_0 = 0.7$, $\Omega_m = 0.3$, and $\Omega_{\Lambda} = 0.7$ throughout this paper.
% COMMENT BY BT

\begin{table*}
    \centering
    \caption{Intrinsic Eddington ratio distribution function in two obscuration bins$^{a}$ }
    \label{tab:_mcmc}
    \begin{tabular}{cccccc}
        \hline
		%Model & $\alpha$ & $\beta$ & $\log ({M}_{\rm BH}^{*})$ & $\delta_1$ & $\epsilon_{\lambda}$ & $\log (\lambda_{*})$ \\ 
		\logNH\ bin & $ {\rm N}_{\rm obs}$$^{b}$ & $\log \xi$ & $\delta_1$ & $\epsilon_{\lambda}$ & $\log (\lambda_{*})$ \\ 
		\hline
		$\logNH < 22$ & 301 (6) & & & & \\
		$\sigma = 0.3$ & & \erdflogxistarunabs & \erdfdeltaaunabs & \erdfepislonlamunabs & \erdfloglamstarunabs \\
		$\sigma = 0.5$ & & \erdflogxistarunabssigfive & \erdfdeltaaunabssigfive & \erdfepislonlamunabssigfive & \erdfloglamstarunabssigfive \\
		$22 \leq \logNH \leq 25$ & 285 (4) & & & \\
		$\sigma = 0.3$& & \erdflogxistarLogNHTwentyTwoToTwentyFive & \erdfdeltaaLogNHTwentyTwoToTwentyFive &  \erdfepislonlamLogNHTwentyTwoToTwentyFive & \erdfloglamstarLogNHTwentyTwoToTwentyFive \\
		$\sigma = 0.5$ & &  \erdflogxistarLogNHTwentyTwoToTwentyFivesigfive & \erdfdeltaaLogNHTwentyTwoToTwentyFivesigfive &  \erdfepislonlamLogNHTwentyTwoToTwentyFivesigfive & \erdfloglamstarLogNHTwentyTwoToTwentyFivesigfive \\
		\hline
    \end{tabular}
    \tablenotetext{a}{Parameters defined in Equation~\ref{eq:erdf_dbpl_def}.}
    \tablenotetext{b}{ Number of AGN with $\log \lamEdd > 0$ in parentheses.}
\end{table*}

\section{Analysis}\label{sec:analysis}

Our main sample follows the same selection criteria as {A22}, and includes all unbeamed AGN over an absolute galactic latitude of $5^{\circ}$, and fall within $0.01\leq z \leq 0.3$, $6.5 \leq \log~({\rm M}_{\rm BH}/{\rm M}_{\odot}) \leq 10.5$ and $-3 \leq \log~\lamEdd \leq 1$ (i.e., a total of 586 sources from the mass, redshift and \lamEdd\ range where the BASS sample is most complete; see Figure~1 in {A22}). We use the masses, redshifts \citep{koss2022_agn_catalog}, X-ray luminosities and obscuration measurements \citep{Ricci2017_Xray_cat} for BASS DR2 sample to calculate Eddington ratios for these AGN.  

\subsection{Intrinsic Eddington Ratio Distribution Functions}

\begin{figure}[ht!]
\includegraphics[width=0.5\textwidth]{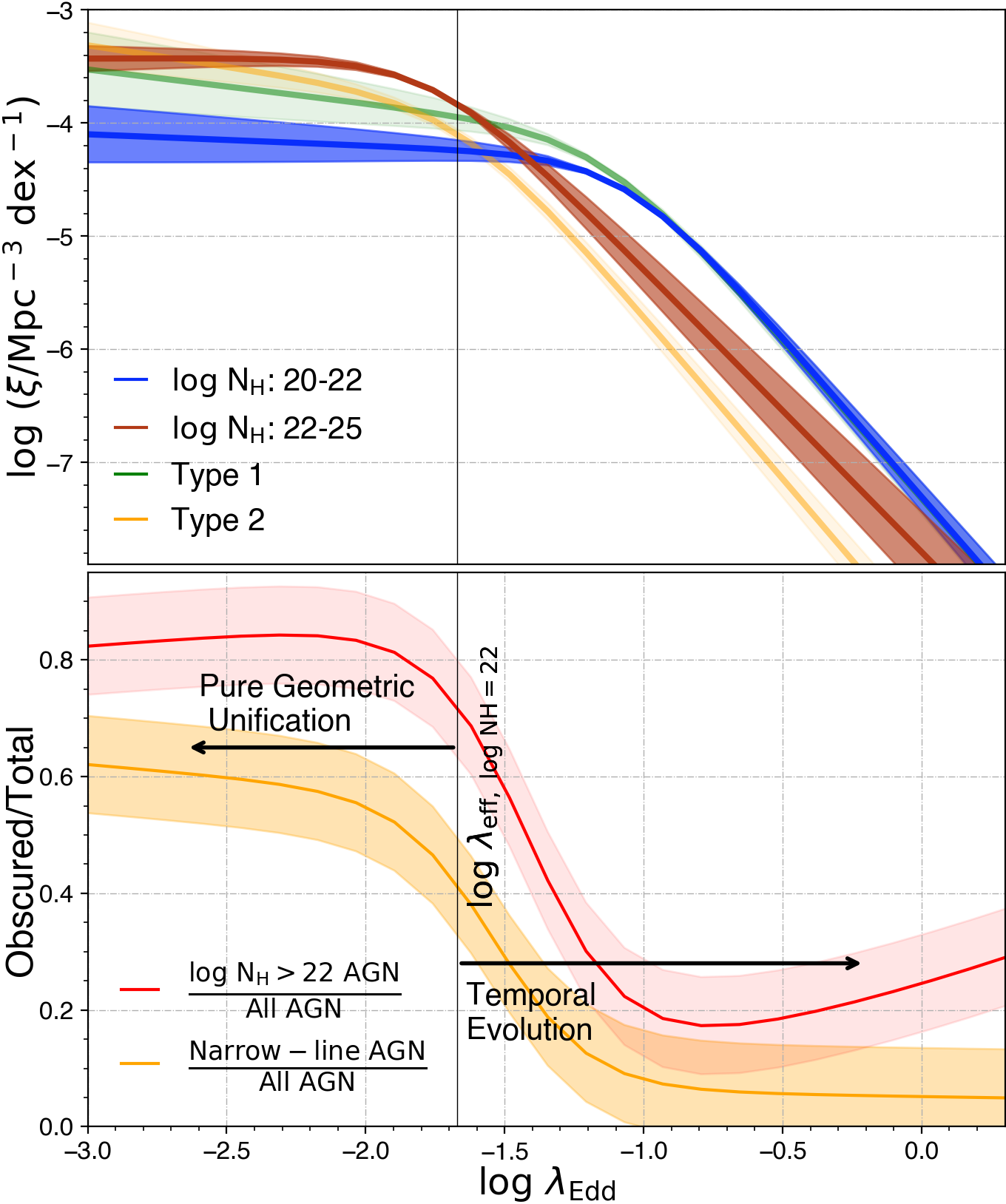}
	\caption{\label{fig:cov_fact_all_bins} {\it Top panel:} Eddington ratio distribution functions for AGN in $\logNH = 20{-}22$ and $\logNH = 22{-}25$ bins, as well as for Type~1 and Type~2 AGN (from {A22}). Shaded regions show 1$\sigma$ uncertainties. % [e.g. $\logNH > 22$]. 
	{\it Bottom panel:} Ratio of obscured to overall space densities as a function of \lamEdd . The \textit{red line} shows ratio of \logNH = 22--25 AGN over all AGN, and the \textit{orange line} shows ratio of Type~2 AGN/all AGN. % as a fraction of all AGN. 
	{\it Black vertical lines} show effective Eddington limits for $\logNH \geq 22$ gas according to \citet{fabian_effect_2008}. At low Eddington ratios (\lamEdd $< -1.7)$, this ratio is high because gas is retained, while for higher Eddington ratios, gas is blown away. We suggest that at low-\lamEdd\ this ratio is equal to the pure geometric covering factor of the torus, while at high \lamEdd , it is a time-averaged covering factor. %Even though AGN with $\log~\lamEdd > -1.67$ should be transitioning according to theoretical models, we choose a conservative cut at a highre $\lamEdd$. 
	We calculate the torus opening angle at low-\lamEdd\ and the covering factor decay rate at high \lamEdd\ using these ratios.}% The dashed vertical lines show the $\lamEdd$ where we quantify the time }% Shaded regions show the 1 $\sigma$ uncertainties.} %The covering factor indicates the presence of broad-line, $\logNH > 22$ AGN at all \lamEdd .  \textcolor{red}{this is covering factor at low lamedd}%, with Compton-thin gas disappearing faster than higher column density gas. 
	%Roughly 25\% of AGN are unobscured at high \lamEdd.
	%} % caption
\end{figure}

%\subsection{Data}\label{sec:data}

%Swift-BAT's high energy X-ray window provided a nearly unbiased sample of local AGN. Over a decade of follow-up multiwavelength and spectroscopic campaigns, the BASS collaboration\footnote{www.bass-survey.com} collected black hole masses for 95\% of the sources in this sample \citep{koss2022_DR2_data,koss2022_DR2_data}. 

While previous studies have constrained the Eddington ratio distribution function for Type~1/broad-line AGN \citep[e.g., ][]{Kelly:2013aa,Schulze:2015aa}, {A22} was the first to constrain the intrinsic Eddington ratio distribution function for obscured Type~2/narrow-line AGN. We briefly describe the method here (more details in \S 3.4 of {A22}). Using a Bayesian ensemble sampler with 50 walkers \citep{emcee}, we maximize the likelihood for the following function:

\begin{equation}\label{eq:likelihood}
    \ln \mathcal{L} =\sum_i^{N_{\rm obs}} \ln p_i(\log M_{\mathrm{BH}, i}, \log N_{\mathrm{H},i}, \log \lambda_{\mathrm{E}, i}, z_i) ,
\end{equation}
where $p_i$ is a convolution of the intrinsic AGN mass function and $\log~\lamEdd$ distribution (constrained together), co-moving volume element, the area-flux curve (i.e., selection function) of the survey(s), redshift evolution function, and absorption function. As $p_i$ is a probability distribution function, it is normalized by integrating this product over all observables:
\begin{equation}\label{eq:bhmfprobconv}
\begin{aligned}
p_{\rm i, conv} & (\log M_{\mathrm{BH}, i}, \log \lambda_{\mathrm{E}, i}, \log N_{\mathrm{H},i}, z_i) = \frac{N_{\rm i, conv}}{N_{\rm tot}}\\ 
& = \frac{1}{N_{\rm tot}}\iiiint \Psi(\log M_{\mathrm{BH}}, \log \lambda_{\mathrm{E}}, \log N_{\mathrm{H}}, z) \\
&\ \times\Omega_{\rm sel}(\log M_{\mathrm{BH}}, \log \lambda_{\mathrm{E}}, \log N_{\rm H}, z)\\
&\ \times p (\log N_{\rm H})~p(z)~\frac{dV_C(z)}{dz}\\
&\ \times \omega(\log M_{\mathrm{BH}, i}, \log \lambda_{\mathrm{E}, i}, \log N_{\mathrm{H}, i}, z_i|\\
&\log \Mbh, \log \lamEdd, \log N_{\mathrm{H}}, z) \\
& ~~d\log\Mbh ~~ d\log\lamEdd ~~ d{\log \NH} ~~ dz \, ,
\end{aligned}
\end{equation}
where $\Psi$ is the convolution of mass and Eddington ratio distribution functions. The parametric form of the intrinsic Eddington ratio distribution function is a double-power law:
\begin{equation}\label{eq:erdf_dbpl_def}
\xi(\log \lamEdd) = \frac{dN}{d\log\lamEdd} \propto \xi^{*} \times \left[\left(\frac{\lamEdd}{\lamEdd^*}\right)^{\delta_1} + \left(\frac{\lamEdd}{\lamEdd^*}\right)^{\delta_2}\right]^{-1} \, .
\end{equation}
$\Omega_{\rm sel}$ is the selection function (area-flux curve for BAT 70-month survey; \citealp{Baumgartner:2013aa}), $p (\log N_{\mathrm{H}, i})$ is the intrinsic absorption function for BAT AGN (from \citealp{Ricci:2015aa}), $p(z_i) = 1$ as we assume negligible redshift evolution over the $0.01 \leq z \leq 0.3$ range, and $\frac{dV_C(z_i)}{dz}$ is the co-moving volume element. The $\omega$ term allows us to convolve uncertainty in mass measurement. We assume Gaussian scatter in mass and luminosity measurement with different dispersions in {A22}. %The simplest case (and main result shown in Figures) assumes $\sigma_{\rm M} = \sigma_{\lamEdd} = 0.3$. %In Appendix D of {A22}, using mock catalogs, we demonstrate that this method correctly recovers the Eddington ratio distribution functions even with $\sigma_{\rm M} = 0.5$ dex. 
%\textbf
{We report results for both $\sigma_{\rm M} = \sigma_{\lamEdd} = 0.3$ and $0.5$ cases in \S~\ref{sec:result_only}, and find that the functions agree within 1$\sigma$ over the range of $\log~\lamEdd$ considered here, so only the first case is shown in the figures for clarity. The 1$\sigma$ random errors on the functions are calculated using the covariance matrix, which we derive from the MCMC chain. The formula for this error estimation is given in Appendix C of {A22}. Note that assuming a functional form such as double-power law leads to smaller errors than more flexible approaches for constraining space densities \citep[e.g.,][]{buchner2015,ananna2020aa}. We assume that the shape of the ERDF is independent of mass, because in \S~4.4 of {A22}, we showed that for the mass range considered for this sample [$6.5 \leq \log~({\rm M}_{\rm BH}/{\rm M}_{\odot}) \leq 10.5$], the shape of the ERDF does not change when constrained in two mass bins independently: [$\log~({\rm M}_{\rm BH}/{\rm M}_{\odot}) \leq 7.8$ and $\log~({\rm M}_{\rm BH}/{\rm M}_{\odot}) \geq 8.2$].}

\subsection{Evolution of AGN Covering factor at high Eddington Ratio}\label{sec:time_decay_method}

%\textbf
{When the torus is stable, the ratio of obscured to overall AGN space density is equal to the population averaged covering factor of the torus. When the radiation pressure is very high (e.g., at high \lamEdd , the torus is unstable, and its geometry is time-dependent). To disentangle the geometric and temporal aspects at high \lamEdd\ (theoretically at $\log~\lamEdd > -1.7$; see \S~\ref{sec:lowlamedd}), we consider two simple parametric models of how the covering factor varies over time at high \lamEdd . We assume that after AGN are triggered from a low-\lamEdd\ phase to high \lamEdd , they typically start with a high covering factor ($\sim 83$\% for $\logNH = 22-25$ torus and $\sim 60\%$ for narrow-line only AGN, as shown in the lower panel of Figure~\ref{fig:cov_fact_all_bins}). As obscuring matter is removed because of the increased radiation pressure around an AGN, its covering factor should decrease. We consider two scenarios for the temporal dependence: (i) the rate at which the covering factor decreases is highest when there is more obscuring matter, or (ii) the covering factor initially decreases slowly due to shielding from obscuring matter, % (see \citealp{fabian2006effedd}), 
and then decreases more rapidly as more and more matter is removed. Both scenarios end with the covering factor decreasing asymptotically such that it \textit{may} approach zero, depending on the obscured/overall ratio.  We do not force the final covering factor to equal zero at the end of the high \lamEdd\ phase as IR studies of luminous quasars show residual dust even for luminous optical quasars \citep[e.g.,][]{hall_2004_dust_redden,hopkins_dust_redden_2004}, essentially allowing the data to decide the final covering factor. We use exponential and sigmoid functions to model the behavior for scenarios (i) and (ii), respectively:} %and normalize to the population-averaged intrinsic covering factors
%(red and yellow lines in bottom panel of Fig. \ref{fig:cov_fact_all_bins}):}
\begin{equation}\label{eqn:cov_fact_time}
\begin{aligned}
&(i)~C_{\rm exp}(t) = C_0 e^{-kt} \\
&(ii)~C_{\rm sigmoid}(t) = C_0 \frac{1+e^{-k t_0}}{1+e^{k(t-t_0)} }
\end{aligned}
\end{equation}

%\textbf
{Figure \ref{fig:decay_function} shows the simulated evolution derived for a sample of 1000 AGN with the average observed covering factors [drawn from a normal distribution of $\mu = 0.83, \sigma = 0.08$ for $\logNH = 22-25$ torus, and $\mu = 0.6, \sigma = 0.07$ for the narrow-line AGN]. Given a set of parameters, we calculate the average covering factor for these 1000 AGN over 100 time steps. %Using an MCMC approach, we find the best-fit parameters that converge on the covering factors shown in Figure~\ref{fig:cov_fact_all_bins} for a given Eddington ratio (over $\log~\lamEdd > -1.67$). %If we were to observe this population of 1000 AGN over the entire length of time spent at this $\lamEdd$ (or a $\lamEdd$ range) at random inclination angles, we should find the %, as
%We let these AGN take 100 steps, varying $k$ and $t_0$ in each step, constrained using an MCMC approach and the covering factor shown in Figure~\ref{fig:cov_fact_all_bins}, averaged over $-1 \leq \log~\lamEdd \leq 0.5$. 
%the fraction of objects in the cleared phase should be equal to the 
To calculate the time-averaged covering factor for these 1000 objects, we draw random times from the lifetime of each AGN, and average over the whole population. We repeat this 100 times for each set of parameters to account for the stochasticity in selecting different time steps, and choose the median. We use a Bayesian ensemble sampler to optimize these functions to reproduce the obscured/overall ratio (shown in Figure~\ref{fig:cov_fact_all_bins}) using a Gaussian likelihood function:}
\begin{equation}
\begin{aligned}
    L(k, t_0;{\rm obs~ratio}) =& \frac{1}{\sqrt{2 \pi \sigma_{\rm obs~ratio}^2}} \times \\ &\exp{\left(-\frac{(\langle C(t) \rangle - \mu_{\rm obs~ratio})^2}{2 \sigma_{\rm obs~ratio}^2}\right)}
\end{aligned}
\end{equation}
where C(t) is defined in Eqn.~\ref{eqn:cov_fact_time}, and only the parameter k is constrained for scenario (i).
%\textcolor{red}{START HERE}Given this population of 1000 objects and a given a set of parameters Using an MCMC approach, we calculate the time-weighted covering factor  , compare it to the time averaged covering factor shown in the `temporal evolution' region of the bottom panel of Figure~\ref{fig:cov_fact_all_bins}. 
%Integrating the median covering factor at each step over the time interval is equal to the overall population covering factor. 

%which leads to the values of $k$ and $t_0$ shown in Figure~\ref{fig:summary_image}. 
%We additionally constrained these functions at $\log~\lamEdd = -1.5$, and found that the covering factors decline more slowly over time and reaches a minimum of 30-40\%. This indicates that as \lamEdd\ increases, the final covering factor decreases (i.e., covering factor will not approach zero at intermedaite \lamEdd\ ).

\section{Results}\label{sec:result_only}

\begin{figure*}[ht!]
    \centering
    \includegraphics[width=.65\textwidth]{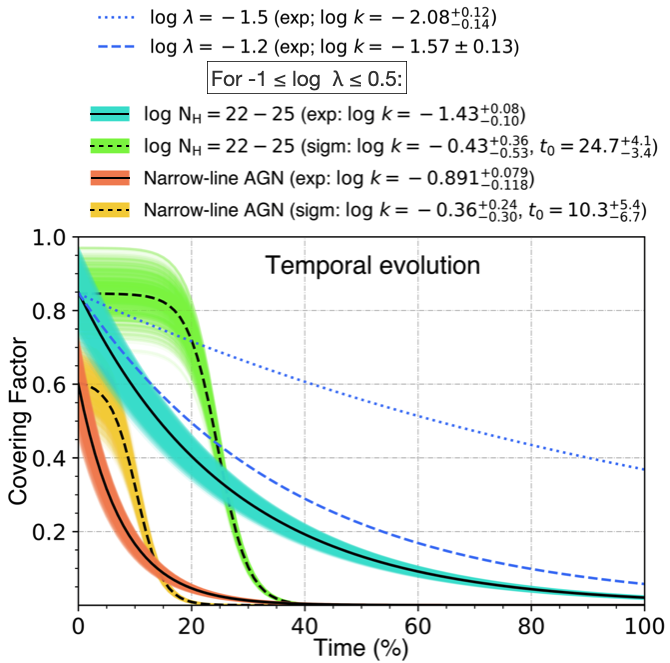}
	\caption{ Decline of the covering factor after an AGN reaches high enough \lamEdd\ that radiation pressure exceeds the gravitational pull on obscuring matter (at $\log~\lamEdd \geq -1.7$; \citealp{fabian_effect_2008}), shown for two different assumed temporal dependences ({\it dashed} and {\it solid black lines} represent sigmoid and exponential functions, respectively, as in Eq. \ref{eqn:cov_fact_time}),
	%; models motivated in \S~\ref{sec:time_decay_method}), 
	evaluated for two different definitions of obscured versus unobscured classes (sorted by $\log~\NH$ or spectral line width). The colored lines show the decay for a 1000 AGN starting from different covering factors (see \S~\ref{sec:time_decay_method}).
	%According to theoretical models \citep[e.g.,][]{fabian_effect_2008}, the effective Eddington limit of obscuring matter with \logNH $ = 22$ is $\log~\lamEdd = -1.67$; above that limit, AGN should transition toward an unobscured state. 
	Assuming that the ratio of obscured to overall AGN 
	%at these Eddington ratios 
	is a time-averaged covering factor (see bottom panel of Figure~\ref{fig:cov_fact_all_bins}), % rather than a single geometric covering factor, 
	we use an ensemble sampler to constrain the model parameters (shown in legend). All \textit{black lines} and associated shaded regions show covering factor decay for $-1 \leq \log~\lamEdd \leq 0.5$.
	%. All the shaded curves listed in the legend are constrained using obscured to overall ratio in the $-1 \leq \log~\lamEdd \leq 0.5$ range. %rate at which the covering factor decays (resulting model parameters shown in legend). 
	Exponential decay functions at two lower Eddington ratios are shown using \textit{blue lines}: \textit{blue dotted lines} show results for $\log~\lamEdd = -1.5$ and \textit{blue dashed lines} show results for $\log~\lamEdd = -1.2$. The residual covering factor is higher at lower \lamEdd . 
	\label{fig:decay_function}}% The exponential scaling factors for these lines are: $\log~k_{-1.5} = -2.08^{+0.12}_{-0.14}$ and $\log~k_{-1.2} = -1.57 \pm 0.13$. } %shown for each obscuration bin in colors corresponding to Fig.~\ref{fig:cov_fact_all_bins})
\end{figure*}

\begin{figure*}[ht!]
    \centering
    \includegraphics[width=.7\textwidth]{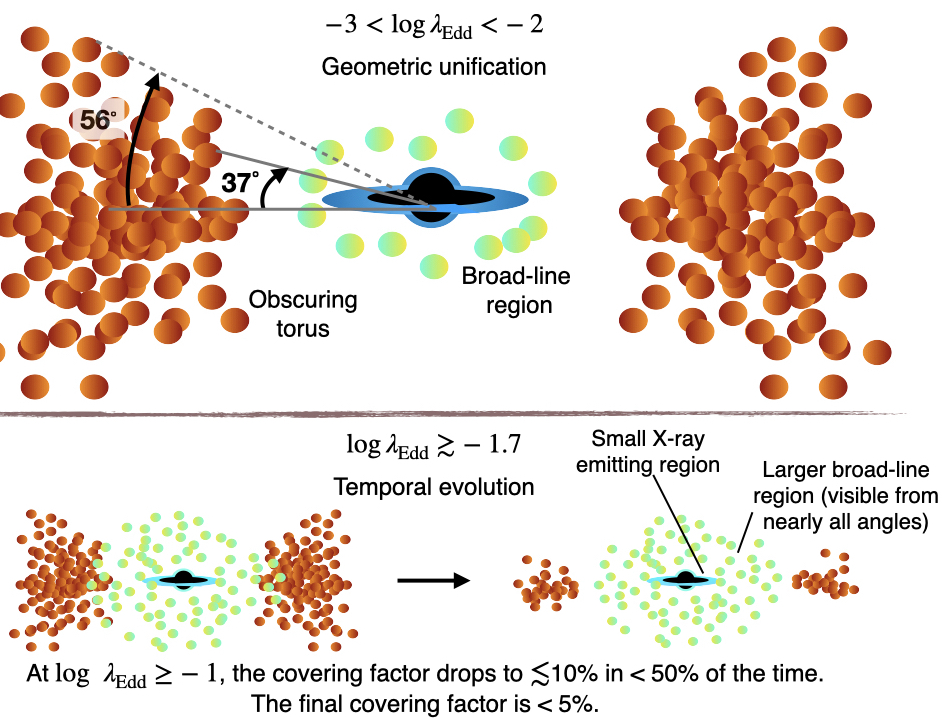}
    \caption{ Schematic diagram of AGN circum-nuclear structure at low and high Eddington ratios.  
    \textit{Top panel:} For $\lamEdd < -2$, radiation pressure is too low to remove dust and gas around the AGN, so the covering factor remains high and differences between obscured and unobscured AGN are likely due to viewing angle. 
    That is, because 80\% of AGN have $\logNH>22$, the torus rises to $56^{\circ}$ above the equatorial plane; since optical broad lines emission are seen in 60\% of AGN, the densest gas lies below 37$^{\circ}$.
    \textit{Bottom panel:} At high \lamEdd , radiation pressure removes the obscuring matter, leaving at most a small persistent covering factor ($< 5\%$). 
    We find that BLR is visible from nearly all angles after 30\% of the time at this \lamEdd , which may indicate that the height of the BLR region rises above that of the depleted torus.
    %At high \lamEdd, clumpy torus blown away, AGN spend 50-70\% of the time with covering factor ($<10$\%).
    \label{fig:summary_image}}
\end{figure*}

% \begin{figure}[ht!]
  %\centering
%    \includegraphics[width=.85\linewidth]{lambda_absorption_function.png}
%    \includegraphics[width=.85\linewidth]{A19_BAT_absorption_functions.png}
	%\includegraphics[width=.95\linewidth]{fraction_of_AGN_over_lamedd}
%	\caption{\textit{Top panel:} The fraction of AGN in each absorption bin {(i.e., area normalized to unity over the five bins)}, for four Eddington ratios. At low \lamEdd , (roughly, at $\log~\lamEdd < -1.67$) these fractions reflect the geometry of the torus and the covering factor in each bin of obscuration. At higher \lamEdd , obscuring matter is likely removed due to radiation pressure, so these fractions indicate the longevity of matter as a function of obscuration. %Note that these absorption functions are normalized to unity over the $\logNH = 20-25$ range.
%	\textit{Bottom panel:} A comparison of the overall $N_{\rm H}$ %\logNH\ 	distribution integrated over $-3 \leq \lamEdd \leq 1$, with the \logNH\ distribution from recent luminosity functions \citep{Ueda:2014aa,ananna2019}, integrated over luminosity ($\log~L_{\RM 2-10} = 42{-}46$), at $z = 0.05$. Note that the overall $N_{\rm H}$ %\logNH\ 	distribution ({\it red line, bottom panel}) is closer to the low \lamEdd\ \logNH\ distribution ({\it red line, top panel}) because the population is dominated by low-\lamEdd\ AGN; there is slightly more than one dex difference in integrated space densities above and below \lamEdd\ $ = -1.5$ (at which point absorbed and unabsorbed space densities are equal).}
%	\vspace{+0.2cm}
%	\label{fig:edd_ratio_absorption_function}
%\end{figure} 

The intrinsic Eddington ratio distribution function in bins of $\log N_{\rm H}$ is presented in Table~\ref{tab:_mcmc} and Figure~\ref{fig:cov_fact_all_bins}. The top panel of Figure~\ref{fig:cov_fact_all_bins} shows the %${\rm N}_{\rm H}$ and $\lamEdd$ distribution for Type~1 and Type~2 AGN within the observed sample, while the bottom panel shows the 
selection bias and 
measurement uncertainty-corrected Eddington ratio distributions of unabsorbed [\logNH : $20{-}22$] and absorbed AGN bins [\logNH : $22{-}25$], along with the Type~1/Type~2 ERDFs from {A22}.  The bottom panel shows the fraction of Type~2 AGN and \logNH : $22{-}25$ AGN as a function of \lamEdd , calculated by dividing the ERDFs of these populations by the ERDF of all AGN. 

{Figure~\ref{fig:decay_function}, shows the evolution of the geometric covering factor at high \lamEdd\ (see Eq.~\ref{eqn:cov_fact_time}).\ %For the functions shown in dashed (sigmoid) and solid (exponential) black lines, we 
These model parameters ($k$ and $t_0$) are constrained using the ratio of obscured to the total population shown in Fig. \ref{fig:cov_fact_all_bins}, at $\log~\lamEdd = -1.5$, -1.2, and averaged over $-1 \leq \log~\lamEdd \leq 0.5$. The average ratios for the last bin are $0.23 \pm 0.08$ and $0.06\pm{+0.07}$, for X-ray and optical measure of obscuration, respectively. %, as the ratio of obscured to overall space density is more stable over this range of $\log~\lamEdd$ (see Fig.~\ref{fig:cov_fact_all_bins}).
%column density and optical categorization, respectively). 
%The best-fit parameters are given in the legend. 
% In Figure~\ref{fig:summary_image}, we offer a simplified schematic of our conclusions, a geometric unification model at low-\lamEdd, and a temporal evolution scenario at high \lamEdd .
}

%The average covering factor for $-3 \leq \log \lamEdd \leq -2$ is 84\% for
%$\logNH = 22{-}25$ AGN. 
% The angles shown in Figure~\ref{fig:summary_image} are calculated accordingly. 
%For comparison, the average covering factors for $-1 \leq \log \lamEdd \leq 0.3$ is 20-30\%. Figure~\ref{fig:summary_image} shows a schematic diagram of circum-nuclear structure at low and high \lamEdd, labeled by the approximate angles subtended by the obscuring material, derived using $\theta = 90^{\circ} - {\rm arccos(covering~factor)}$. 
%the angles corresponding to this covering factors are labeled. The in the low \lamEdd\ case in the figure. 
%For simplicity, we show the obscuring material as a solid torus in the figure, although it is likely the distribution of matter is clumpy rather than uniform for each column density \citep[e.g.,][]{almeida2009,stalevski2012,elitzur_unification_2012, mislav2018,zhao_properties_2021}.

\section{Discussion}\label{sec:discussion_only}

%\subsection{The Torus and the Broad-line region}

%\textbf
{ In this work, we calculate the Eddington ratio distribution functions for obscured and unobscured AGN using X-ray measures of \NH . We compare these functions to the ERDF for Type~1 and Type~2 AGN calculated in {A22} using the same sample (all ERDFs shown in Fig. \ref{fig:cov_fact_all_bins}).}

%\textbf
{At low Eddington ratio, when the circum-nuclear torus is presumably stable, the ratio of the space density of obscured AGN to all AGN reflects the average covering factor of the torus. In \S~\ref{sec:lowlamedd}, we discuss the torus at low-\lamEdd , below the theoretical threshold where radiation pressure exceeds the gravitational pull. 
At high \lamEdd, in contrast, the gas and dust are blown away, so this ratio reflects the time-averaged covering factor. In \S~\ref{sec:highlamedd}, we consider the ERDF and obscured/overall ratio at higher \lamEdd, and explore the time dependence of the covering factor using simple physically motivated models described in \S~\ref{sec:time_decay_method}.}%As the broad-line region (BLR) in AGN are more extended than the X-ray emitting coronal region, it is sampled by more lines of sight, which in the case of a clumpy geometry, means AGN with broad lines may still have high \NH\ along our line of sight. From the top panel of Figure~\ref{fig:2d_scatter_plot}, we see that nearly all purely narrow-line AGNs are at $\logNH > 22$, as expected. 
%At the same time, roughly a quarter of broad-line AGN also have high \NH , which is why the Type~2 ERDF lies below the $\logNH > 22$ ERDF (Fig.~\ref{fig:cov_fact_all_bins}). 
%The detailed physical implications of these findings are discussed below. 

%the Eddington ratio at which gas is blown out, and the stability of the torus at lower-\lamEdd\ is discussed in \S~\ref{sec:lowlamedd}. I

% From the topmost panel of Figure~\ref{fig:2d_scatter_plot}, we see that nearly all purely narrow-line AGN (with no broad-line detection) are at $\logNH > 22$. In the lower panel of Figure~\ref{fig:cov_fact_all_bins}, this implies that all narrow-line AGN are contained within the $22 < \logNH < 25$ covering factor (i.e., the objects represented by the orange line is a subset of objects represented by the red line). %These leaves $\simeq 1520$\% of AGN at nearly which are detected as both broad-line and obscured in X-rays.

\subsection{Low-\lamEdd\ Region: Geometric Unification}\label{sec:lowlamedd}

\citet{fabian2006effedd,fabian_effect_2008} pointed out that the standard Eddington ratio is calculated for fully ionized gas, whereas for partially ionized cold gas, the scattering cross-section is higher, and therefore the ``effective Eddington limit'' (\lamEddeff) for such gas is lower.  That is, colder, less ionized gas can be removed from near the AGN at lower levels of radiation pressure and thus lower Eddington limit.

% Here we adopt a simplistic assumption to compare our results quantitatively with this theoretical model: assuming 
In order to compare our results quantitatively with this theoretical picture, we check whether
$\lambda^{*}$, the break in the power-law form of the Eddington ratio distribution function (see Eqn.~\ref{eq:erdf_dbpl_def}), %for each column density bin, 
corresponds to the effective Eddington ratio for dusty gas at high densities. 
The value of $A$ from Fig. 1 in \citet{fabian_effect_2008} represents the ratio of scattering cross sections for dusty gas compared to ionized gas (see Eqn. 2 from \citealp{fabian_effect_2008}), or equivalently, to \lamEdd/\lamEddeff. 
That is, the break, $\lambda^{*}$, should occur at $\lamEddeff/A$ (for $\lamEdd=1$). 
%calculate the multiplicative factor A($\log N_{\rm H}$), $\log~A = -\log \lambda_{*}$ \citep{fabian2006effedd}, using which the scattering cross-section can be calculated trivially. 
%\textbf
{For $\logNH = 22{-}25$, $\log\lambda^{*} =$ \erdfloglamstarLogNHTwentyTwoToTwentyFivesigfive\ for $\sigma_{\log~M} = \sigma_{\log~\lambda} = 0.5$ , which is in excellent agreement with the \citet{fabian_effect_2008} predicted value, $\log A \sim1.67$. }

Under \lamEddeff\ for obscured gas, both theoretically and as indicated by observations ($\log~\lamEdd < -1.67$ according to \citealp{fabian_effect_2008} and our results, and $\log~\lamEdd < -2$ more conservatively), the radiation pressure is too low to blow away the obscuring matter. The simple scenario that emerges from this low-\lamEdd\ region is that the pure geometric unification model \citep{antonucci1993,urry1995,netzer2015} applies in this $\lamEdd$ regime, where the covering factor of the torus (equal to the ratio of obscured to all AGN in this regime $\lamEdd$) could be as high as 83\%. However, the BLR is completely blocked for only 60\% of the total solid angle, which could be indicative of clumpiness in the torus structure \citep[e.g.,][]{almeida2009,stalevski2012,elitzur_unification_2012} allowing some broad-line visibility even with obscuring matter along our line of sight.

Semi-analytical models \citep[e.g.,][]{venanzi_role_2020} find that the densest gas sinks closer to the equatorial plane of the AGN, due to asymmetry of the radiation field ($\propto \cos{\theta_{\rm axis}}$). If the pure geometric unification model applies, this would mean that our view of the BLR is blocked when looking through angles closer to the equatorial region, whereas at angles higher above the equatorial plane, obscuring matter is distributed more sparsely, and therefore the BLR is more likely to be visible. According to our calculations with the BASS sample, at $-3 < \log~\lamEdd < -2$, the torus rises as high as $56^{\circ}$ [calculated using $\theta = 90^{\circ} - {\rm arccos(covering~factor)}$; similarly, the torus opening angle $ = {\rm arccos(covering~factor)} = 34^{\circ}$], while the BLR is completely blocked by dense matter up to $37^{\circ}$ above the equatorial plane. %\textbf
{A schematic of the geometric unification model, along with these angles and the clumpy structure of the torus, is shown in the top panel of Figure~\ref{fig:summary_image}. } %These angles are calculated using this relationship: . %, and the angles are given in the bottom panel of Figure~\ref{fig:angles} as a function of \lamEdd .}

%The angles for broad-line isn't accurate, because it can only be a lower limit, but the torus reflects pure geometric unification.

\subsection{High-\lamEdd\ Region: Transitional Timescales}\label{sec:highlamedd}

%\textcolor{red}{The angles for broad-line region is the actual covering factor of broad-line AGN. The obscure 'broad-line' AGN are HotDOGs and red-quasars observed at higer redshifts. %At higher redshifts there are more of them, possibly because more dust/gas was available for obscuration. 
%We find a low fraction of this population ($\simeq$ 15-20\%) - exactly in line with glikman papers.}

The high-\lamEdd\ region gives rise to some seemingly contradictory observational signatures. While overall population studies such as ours indicate that the covering factor at high \lamEdd\ should be very low if geometric unification applies (e.g., lower panel of Figure~\ref{fig:cov_fact_all_bins}, Figure 3 of \citealp{Ricci2017Nat}),  \textit{observed} obscured AGN which are found at these \lamEdd\ have very high covering factors. \citet{Ricci2017_mergers} finds that among local ultra-luminous infrared galaxies (ULIRGs), the torus covering factor is as high as 95\%, and most of these AGN are in late stages of mergers. A recent study of 57 ULIRGs \citep{yamada_comprehensive_2021} also finds a high covering factor of 66\% at $\lamEdd \simeq 1$. {This is quantitatively at odds with a population-averaged covering factor of $<30$\% at these \lamEdd\ (from Fig.~\ref{fig:cov_fact_all_bins}). 
While ULIRGs represent only a small fraction of the overall AGN population, they dominate infrared samples because they have a lot of dust and they have high covering factors and high \lamEdd .
Other obscured high-$\lamEdd$ populations include red quasars \citep[e.g.,][]{Glikman2004,Glikman2012,banerji_heavily_2015,Glikman2018}, and Hot Dust-Obscured Galaxies \citep[HotDOGs, e.g.,][]{assef_hot_2016,vito_heavy_2018}. 
%Red quasar samples are at relatively lower redshifts and are less obscured than HotDOGs, 
%Red quasars are less obscured than HotDOGs and thus may be a transitional phase between HotDOGs and optical quasars \citep[e.g.,][]{assef_hot_2016,vito_heavy_2018,wu_eddington-limited_2018,jun_spectral_2020}.
Some X-ray selected studies also find that heavily dust-reddened %`forbidden region' 
quasars %at $z > 2$ 
are in a radiatively driven blow-out phase \citep{lansbury_x-ray_2020}.}

%\textbf
{In other words, infrared-selected quasars have high covering factors and high \lamEdd . Yet, according to Figure~\ref{fig:cov_fact_all_bins}, we should find a much smaller obscured/overall ratio, implying a much smaller torus covering factor if simple geometric unification were the cause. 
%\citet{yamada_comprehensive_2021}
%\textbf{These IR selected samples show that these covering factors can be very high.} 
%Regardless of how these AGN are triggered, we note the tension between population studies finding low covering factors and individual studies of obscured luminous AGN finding higher factors. 
As most of these studies conclude that these high-\lamEdd, obscured AGN are in a transitional state, we suggest that the low obscured/overall ratio at high \lamEdd\ is indicative of the duration of the obscured phase.
That is, the time-averaged covering factor is low ($\sim30$\%), while infrared studies select luminous/high-\lamEdd\ sources that are still in a dust-obscured phase. %, as the dust re-radiates in infrared.
}

%\textbf
{To disentangle the geometric and temporal aspects of the evolution of the covering factor, we consider two simple time-dependent parametric models of evolution, described in \S~\ref{sec:time_decay_method}. 
In Figure~\ref{fig:decay_function}, we show these functions constrained using the obscured/overall ratio (from Figure~\ref{fig:cov_fact_all_bins}) at $\log~\lamEdd=-1.5$ and $-1.2$, and over the $-1 \leq \log~\lamEdd \leq 0.5$ range. 
%For the former, we consider the average ratio as this value remains approximately constant over this range. 
% In Figure~\ref{fig:decay_function}, we show these functions constrained using the obscured/overall ratio (from Figure~\ref{fig:cov_fact_all_bins}) at three $\log~\lamEdd$ values: -1.5, -1.2, and in the $-1 \leq \log~\lamEdd \leq 0.5$ bin. In the last bin, we consider the average ratio as this value remains approximately constant over this range. 
%We conservatively choose $\log~\lamEdd > -1$ as the temporal evolution phase, however from the \citet{fabian_effect_2008} models, $\log~\lamEdd \simeq -1.67$ is the Eddington limit for $\logNH > 22$ matter, therefore we also show the covering factor decay at $\log~\lamEdd = -1.5$ and -1.2, assuming the exponential model. 
We find that for the first model (the sigmoid function), in which the decay of the covering factor might start slowly due to shielding from dense obscuring matter, then accelerate as more matter is removed, the covering factor for the torus approaches zero within 40\% of the time spent at $\log~\lamEdd > -1$. The sigmoid function therefore somewhat contradicts infrared studies that find evidence of residual dust around luminous SDSS quasars. The second model (exponential function) accommodates slower decay of the X-ray detected covering factor, even though the part of the torus that completely blocks the BLR is depleted inwithin 30\% of the AGN lifetime. %This reveals that enough obscuring matter is removed even from the dense equatorial region so that the BLR becomes visible from essentially all angles. 
The \textit{persistent} covering factors, calculated using X-ray obscuration measurements, at $\log~\lamEdd = -1.5$, $-1.2$ and $\sim 0$ are $\sim 40\%$, 5\% and 2\%, respectively, indicate that at lower $\lamEdd$, more obscuring matter remains around the AGN at the end of the active phase. The bottom panels of Figure~\ref{fig:summary_image} shows a schematic of this transition, with the residual dust ($< 5\%$ by the end of the phase) in the right panel. }

% in an active state encompassing all the flickering episodes, 
%summing over all flickering active states, 

%\textbf{Within 30\% of the time, the covering factor of the overall torus (now with enough clumps removed so that it is entirely broad-line) falls to 30\% (i.e., $ \theta_{\rm opening~angle} \simeq 72^{\circ}$) for both parametric functions. %If we look at the sigmoid function, which allows a sharp transition point for geometric covering factor, we see that the transition occurs at time step $t_0 = 24.7^{+4.1}_{-2.4}$. 
%as relative timescales, we find that AGN spend about 20--30\% of the time in \logNH = 22--25 bins, and 15--25\% in broad-line and \logNH = 22--25 state. This

%\textbf
{While the exponential function allows a gradual decrease in covering factor, the sigmoid function behaves almost like a step function. If we interpret the sigmoid function as a limiting case where the high-\lamEdd\ obscured/total ratio is decided purely by timescale (i.e., an obscured/total ratio of $x$\% means $x$\% of the time is spent fully obscured, and ($100-x$)\% of the time is spent in the almost fully unobscured state), then the percentage of time spent fully obscured by $\logNH = 22{-}25$ torus is $24.7^{+4.1}_{-3.4}$\%. The narrow-line only covering factor will be depleted within $10.3^{+5.4}_{-6.7}$\% of the time, so that for about 15\% of the time the AGN will appear broad-line and obscured. This estimate agrees well with the \citet{Glikman2012} and \citet{Glikman2018} results, which suggested that the duration of the red-quasar phase is 15--20\% of the total quasar lifetime.}

{As both Figures~\ref{fig:cov_fact_all_bins} and \ref{fig:decay_function} show, the X-ray measure of obscured/overall ratio at high-\lamEdd\ is higher than the Type~2/overall ratio. A likely interpretation for this is that the former indicates a time-averaged covering factor, while the latter indicates an increase in the size of the broad-line region relative to the torus. When the scale height of the BLR region is higher than that of the torus, the BLR should be visible even from equatorial lines of sight, as shown in the bottom right panel of Figure~\ref{fig:summary_image}. This depletion of torus/rise of the BLR happens within 30\% of the time spent at $\log~\lamEdd > -1$ (see Fig.~\ref{fig:decay_function}). We do an order-of-magnitude calculation %of time spent in different phases of obscuration 
using well-established constraints on the duration of unobscured quasar accretion, which is roughly the Salpeter time of $10^7{-}10^8$ years \citep{martini_qso_2003,Hopkins2006_QLF_model,worseck_transverse_2007,goncalves_detection_2008}. Using the functions shown in Figure~\ref{fig:decay_function}, we estimate that after being triggered to $\log~\lamEdd > -1$, it takes 20--30\% of the time (3--30 Myr) to transition into a completely broad-line phase. }

\section{Conclusions}\label{sec:conclusion}
%\textcolor{red}{start here}

{This work presents the first \textit{intrinsic} Eddington ratio distribution functions for X-ray-obscured and -unobscured AGN, constrained using the local BASS sample.
%($\logNH > 22$) ($\logNH <22$)
These ERDFs show that there are $\sim 4$ times as many obscured as unobscured AGN at low \lamEdd , while the reverse is true at high \lamEdd , with $\sim 3$ times as many unobscured as obscured AGN.}

{Reasoning that the circum-nuclear obscuration is relatively stable at $\log~\lamEdd < -2$, we interpret that population ratio in purely geometric terms. A ratio of 4:1 corresponds to an $\sim80$\% covering factor, meaning that a simple obscuring torus would rise 56$^{\circ}$ above the equatorial plane. The Type~2 to overall AGN space density ratio,
determined by {A22},
is somewhat smaller, $\sim60$\%, with broad optical lines seen in roughly one quarter of high-\NH\ AGN.
This suggests that the obscuring torus is clumpy. Recent simulations suggest that obscuration is densest near the equatorial plane \citep[e.g., ][]{venanzi_role_2020}. In that case, all broad-lines are blocked within 37$^{\circ}$ of the plane by tightly packed obscuring matter, with some open lines of sight between 37$^{\circ}$ and 56$^{\circ}$.
The top panel of Figure~\ref{fig:summary_image} shows a schematic representation of this clumpy torus, with geometry derived from the ERDF ratios. }

{At high Eddington ratios, the fraction of obscured AGN is much smaller. Additionally, the obscured fraction is much lower for narrow-line AGN ($\sim5$\%) than for AGN with $\logNH >22$ %($\sim30$\%).
($\sim30$\%, see bottom panel of Fig.~\ref{fig:cov_fact_all_bins}). Infrared-selected luminous, high-\lamEdd , obscured AGN have very high covering factors (66-95\%; see \S~\ref{sec:highlamedd}), 
---much  higher than the observed population average.
%which seemingly contradicts our result at high \lamEdd .
%At the same time, obscured AGN in this \lamEdd\ regime have very high covering factors (see \S~\ref{sec:highlamedd}).
This tension can be resolved if the highly covered phase is short-lived and infrared-luminous, so that infrared selection preferentially finds ULIRGs, red quasars, and HotDOGs.
% and infrared selection only chooses AGN in this phase (in the form of ULIRGs/red quasars/HotDOGs). 
}

{To disentangle the geometric and temporal aspects at high \lamEdd , %in \S~\ref{sec:time_decay_method}, we describe 
we considered two simple physically motivated models of the decline in covering factor with time (Fig.~\ref{fig:decay_function}), constrained by the obscured/overall ratios at high \lamEdd . %(from Fig.~\ref{fig:cov_fact_all_bins}). 
Using these models, %(Fig~\ref{fig:decay_function}), 
we find that 
%there is a persistent torus with covering factor near 5\%---these AGN would have both high \NH\ and broad lines---
it takes approximately 50\% of the lifetime of the high-\lamEdd\ phase (i.e., at $\log~\lamEdd > -1$) to reduce the covering factor from 80\% to $\sim$ 10\%. The covering factor is $< 5\%$ at the end of the high-\lamEdd\ phase. Additionally, %the dense equatorial gas is depleted rapidly, dropping to less than 5\% in 20\% of the time
the broad-line region becomes visible along all lines of sight within 20--30\% of the time (orange and yellow lines in Fig. \ref{fig:decay_function}), possibly indicating that the height of the broad-line region is higher than that of the obscuring torus, making it visible along all lines of sight, as shown in the bottom right panel of Figure~\ref{fig:summary_image}.}%At that point, broad lines are visible in all transitioning AGN.
\section*{Acknowledgments}
%+++++++++++++++++++++++++++++++++

We thank the referee for thoughtful comments that helped us greatly improve this paper. This paper is part of a series presented by the BASS Collaboration. Specifically, this is BASS XXXVIII. T.T.A. support from NASA through ADAP award NNH22ZDA001N. T.T.A. and R.C.H. acknowledge support from NASA through ADAP award 80NSSC19K0580, and the National Science Foundation through CAREER award 1554584. C.M.U. acknowledges support from the National Science Foundation under Grant No. AST-1715512, and from NASA through ADAP award 80NSSC18K0418. P.N. acknowledges the Black Hole Initiative (BHI) at Harvard University, which is supported by grants from the Gordon and Betty Moore Foundation and the John Templeton Foundation. B.T. acknowledges support from the Israel Science Foundation (grant number 1849/19) and from the European Research Council (ERC) under the European Union's Horizon 2020 research and innovation program (grant agreement number 950533). M.K. acknowledges support from NASA through ADAP award NNH16CT03C. M.B. acknowledges support from the YCAA Prize Postdoctoral Fellowship. K.O. acknowledges the support of the Korea Astronomy and Space Science Institute under the R\&D program (Project No. 2022-1-830-06) supervised by the Ministry of Science and ICT and from the National Research Foundation of Korea (NRF-2020R1C1C1005462). We acknowledge funding support through ANID programs: Millennium Science Initiative NCN19\_058 (E.T.), and ICN12\_009 (FEB); CATA-BASAL - ACE210002 (E.T., F.E.B.) and FB210003 (E.T., F.E.B., C.R.); FONDECYT Regular - 1190818 (E.T., F.E.B.) and 1200495 (F.E.B., E.T.); and FONDECYT Iniciacion 11190831 (C.R.). This work was performed in part at the Aspen Center for Physics, which is supported by National Science Foundation grant PHY-1607611.

\bibliography{sample631}{}
\bibliographystyle{aasjournal}

%% the collaboration and author truncation commands are used. It has to


\end{document}